\documentclass[12pt]{article}
\usepackage{latexsym,amsmath}
\input amssym
\begin{document}
\renewcommand{\theequation}{\thesection.\arabic{equation}}
\def\prg#1{\medskip{\bf #1}}
\def\lra{\leftrightarrow}        \def\Ra{\Rightarrow}
\def\nin{\noindent}              \def\pd{\partial}
\def\dis{\displaystyle}          \def\dfrac{\dis\frac}
\def\grl{{GR$_\Lambda$}}         \def\vsm{\vspace{-10pt}}
\def\Lra{{\Leftrightarrow}}      \def\ads3{AdS$_3$}
\def\cs{{\scriptscriptstyle \rm CS}}  \def\ads3{{\rm AdS$_3$}}
\def\Leff{\hbox{$\mit\L_{\hspace{.6pt}\rm eff}\,$}}
\def\Tr{\hbox{\rm Tr\hspace{1pt}}}
\def\gp{($\Bbb{P}$)}

\def\D{{\Delta}}      \def\bC{{\bar C}}     \def\bT{{\bar T}}
\def\bH{{\bar H}}     \def\bL{{\bar L}}     \def\bI{{\bar I}}
\def\hO{{\hat O}}     \def\hG{{\hat G}}     \def\tG{{\tilde G}}
\def\cL{{\cal L}}     \def\cM{{\cal M }}    \def\cE{{\cal E}}
\def\cA{{\cal A}}     \def\cI{{\cal I}}     \def\cC{{\cal C}}
\def\cF{{\cal F}}     \def\hcF{\hat{\cF}}   \def\bcF{{\bar\cF}}
\def\cH{{\cal H}}     \def\hcH{\hat{\cH}}   \def\bcH{{\bar\cH}}
\def\cK{{\cal K}}     \def\hcK{\hat{\cK}}   \def\bcK{{\bar\cK}}
\def\cO{{\cal O}}     \def\hcO{\hat{\cal O}} \def\tR{{\tilde R}}

\def\G{\Gamma}        \def\S{\Sigma}        \def\L{{\mit\Lambda}}
\def\a{\alpha}        \def\b{\beta}         \def\g{\gamma}
\def\d{\delta}        \def\m{\mu}           \def\n{\nu}
\def\th{\theta}       \def\k{\kappa}        \def\l{\lambda}
\def\vphi{\varphi}    \def\ve{\varepsilon}  \def\p{\pi}
\def\r{\rho}          \def\Om{\Omega}       \def\om{\omega}
\def\s{\sigma}        \def\t{\tau}          \def\eps{\epsilon}
\def\ups{\upsilon}    \def\tom{{\tilde\om}} \def\bw{{\bar w}}
\def\nn{\nonumber}
\def\be{\begin{equation}}             \def\ee{\end{equation}}
\def\ba#1{\begin{array}{#1}}          \def\ea{\end{array}}
\def\bea{\begin{eqnarray} }           \def\eea{\end{eqnarray} }
\def\beann{\begin{eqnarray*} }        \def\eeann{\end{eqnarray*} }
\def\beal{\begin{eqalign}}            \def\eeal{\end{eqalign}}
\def\lab#1{\label{eq:#1}}             \def\eq#1{(\ref{eq:#1})}
\def\bsubeq{\begin{subequations}}     \def\esubeq{\end{subequations}}
\def\bitem{\begin{itemize}}           \def\eitem{\end{itemize}}

\title{Stability of 3D black hole with torsion}

\author{B. Cvetkovi\'c and M. Blagojevi\'c
\footnote{Email addresses: {\tt cbranislav@phy.bg.ac.yu,
                                mb@phy.bg.ac.yu}} \\
Institute of Physics, P. O. Box 57, 11001 Belgrade, Serbia}
\date{}
\maketitle
\begin{abstract}
Using $N=1+1$ supersymmetric extension of the three-dimensional
gravity with torsion, we show that a generic black hole has no
exact supersymmetries, the extremal black hole has only one, while
the zero-energy black hole has two. Combining these results with
the asymptotic supersymmetry algebra, we are naturally led to
interpret the zero-energy black hole as the ground state of the
Ramond sector, and analogously, the anti-de Sitter solution as the
ground state of the Neveau-Schwartz sector.\\ \\
{\it PACS Number(s): 04.50+h, 04.60.Kz, 04.65+e, 04.70.Bw}  
\end{abstract}

\section{Introduction}
\setcounter{equation}{0}

In the previous two and a half decades, three-dimensional (3D)
gravity has been successfully used for exploring basic features of
the gravitational dynamics. Within the traditional approach based
on general relativity (GR) and the underlying {\it Riemannian\/}
geometry of spacetime, a number of remarkable results has been
achieved \cite{1,2,3,4,x1,5}. However, 3D gravity also represents
an arena for testing a more general, gauge-theoretic conception of
gravity based on {\it Riemann-Cartan\/} geometry---the geometry
that is characterized by both the {\it curvature\/} and the {\it
torsion\/} \cite{6,x2}. Today, fifteen years after Mielke and Baekler
proposed a general topological model for 3D gravity with torsion
\cite{7,x3}, this approach represents a respectable framework for
studying the gravitational dynamics \cite{8,x4,9,x5,10,x6,11,x7,x8,12,13,14}.

Recently, using $N=1+1$ supersymmetric extension of 3D gravity with
torsion, it was shown that there exists a suitable supersymmetric
generalization of the anti-de Sitter (AdS) asymptotic conditions,
which leads to the asymptotic superconformal symmetry \cite{13}. In
the present paper, we explore exact supersymmetries of the black hole
with torsion and use the asymptotic supersymmetry algebra to study
its stability properties. Our results are a natural generalization of
those obtained for the BTZ black hole in \cite{15,x9}.

The paper is organized as follows. In section 2 we recall some basic
aspects of the topological 3D gravity with torsion and its
supersymmetric extension with $N=1+1$ gravitini. Sections 3 and 4
contain basic results of the paper. In section 3, we solve the
Killing spinor equation for the general black hole configuration and
show that: (1) a generic black hole has no exact supersymmetries, (2)
the extremal black hole has only one, while (3) the black hole with
zero energy has two exact supersymmetries. In section 4, we combine
these results with the asymptotic supersymmetry algebra to show that
the zero-energy black hole can be interpreted as the ground state of
the Ramond sector, with periodic boundary conditions for gravitini.
Similarly, the AdS solution is identified as the ground state of the
Neveau-Schwartz sector, with anti-periodic boundary conditions.
Finally, section 5 is devoted to concluding remarks.

Our conventions are the same as in \cite{13}: the Latin indices
$(i,j,k,\dots)$ refer to the local orthonormal frame, the Greek
indices $(\m,\n,\r,\dots)$ refer to the coordinate frame, and both
run over $0,1,2$; the metric components in the local Lorentz frame
are $\eta_{ij}=(+,-,-)$; totally antisymmetric object $\ve^{ijk}$ is
normalized by $\ve^{012}=+1$; gamma matrices are pure imaginary,
$(\g_0,\g_1,\g_2)=(-\s^2,i\s^3,i\s^1)$ with $\s^k$ the Pauli
matrices, and Majorana spinors are real.

\section{Supersymmetric 3D gravity with torsion} 
\setcounter{equation}{0}

\prg{Riemann-Cartan geometry.} Theory of gravity with torsion can be
naturally described as Poincar\'e gauge theory (PGT), with an
underlying spacetime structure corresponding to Riemann-Cartan
geometry \cite{6,x2}. Basic gravitational variables in PGT are the triad
field $b^i$ and the Lorentz connection $A^{ij}=-A^{ji}$ (1-forms),
and the corresponding field strengths are the torsion $T^i$ and the
curvature $R^{ij}$ (2-forms). In 3D, we can simplify the notation by
introducing $A^{ij}=:-\ve^{ij}{_k}\om^k$ and
$R^{ij}=:-\ve^{ij}{_k}R^k$, which yields:
\be
T^i=db^i+\ve^i{}_{jk}\om^j\wedge b^k \, ,\qquad
R^i=d\om^i+\frac{1}{2}\,\ve^i{}_{jk}\om^j\wedge\om^k\, .   \lab{2.1}
\ee

Gauge symmetries of the theory are local translations and local
Lorentz rotations. The covariant derivative $\nabla\equiv\nabla(\om)$
acts on a general tangent-frame spinor/tensor in accordance with its
spinorial/tensorial structure; when $X$ is a form, $\nabla
X:=\nabla\wedge X$.

The metric structure of PGT is defined by $g=\eta_{ij}b^i\otimes
b^j$. Metric and connection are related to each other by the {\it
metricity condition\/}, $\nabla g=0$, which corresponds to {\it
Riemann-Cartan geometry\/} of spacetime. In PGT, we have a useful
identity
\be
\om^i\equiv\tom^i+K^i\, ,                                  \lab{2.2}
\ee
where $\tom^i$ is the Levi-Civita (Riemannian) connection, and $K^i$
is the contortion 1-form, defined implicitly by
$T^i=\ve^i{}_{mn}K^m\wedge b^n$.

\prg{Generalized dynamics.} General gravitational dynamics in
Riemann-Cartan spacetime is determined by Lagrangians which are at most
quadratic in field strengths. Omitting the quadratic terms, we arrive
at the {\it topological\/} Mielke-Baekler model for 3D gravity
\cite{7,x3}:
\be
I_0=2a\int b^i R_i-\frac{\L}{3}\,\int\ve_{ijk}b^i b^j b^k\
    +\a_3I_\cs[\om]+\a_4 \int b^i T_i\, ,                  \lab{2.3}
\ee
where the wedge product sign $\wedge$ is omitted for simplicity. The
first term with $a=1/16\pi G$ is the usual Einstein-Cartan action, the
second term is a cosmological term, $I_\cs[\om]$ is the Chern-Simons
action for the Lorentz connection,
$I_\cs[\om]=\int\left(\om^i d\om_i
           +\frac{1}{3}\ve_{ijk}\om^i\om^j\om^k\right)$,
and the last term is a torsion counterpart of the first one. The
Mielke-Baekler model is a natural generalization of GR with a
cosmological constant.

In the sector $\a_3\a_4-a^2\ne 0$, the vacuum field equations take
the simple form
\be
2T^i=p\ve^i{}_{jk}\,b^j b^k\, ,\qquad
2R^i=q\ve^i{}_{jk}\,b^j b^k\, ,                            \lab{2.4}
\ee
where
$$
p=\frac{\a_3\L+\a_4 a}{\a_3\a_4-a^2}\, ,\qquad
q=-\frac{(\a_4)^2+a\L}{\a_3\a_4-a^2}\, .
$$
Thus, vacuum solutions are characterized by constant torsion and
constant curvature.

In Riemann-Cartan spacetime, one can use the identity \eq{2.2} to
express the curvature $R^i=R^i(\om)$ in terms of its {\it
Riemannian\/} piece $\tR^i=R^i(\tom)$ and the contortion. The
resulting identity, combined with the {\it on-shell\/} relation
$K^i=p\,b^i/2$, leads to
\be
\tR^{ij}=-\Leff\,b^i\wedge b^j\, ,\qquad
\Leff:= q-\frac{1}{4}p^2\, ,                               \lab{2.5}
\ee
where $\Leff$ is the effective cosmological constant. The form of
$\tR^{ij}$ implies that our spacetime has maximally symmetric metric.

\prg{Black hole with torsion.} For negative $\Leff$, the
Mielke-Baekler model has an exact vacuum solution, the black hole
with torsion \cite{8,x4,9,x5,10,x6}, which is a natural generalization of
the well-known BTZ black hole \cite{4,x1}. In static coordinates
$x^\m=(t,r,\vphi)$ (with $0\le\vphi<2\pi$), the BTZ metric is
given as \bea
&&ds^2=N^2dt^2-N^{-2}dr^2-r^2(d\vphi+N_\vphi dt)^2\, ,     \nn\\
&&N^2=\left(-8Gm+\frac{r^2}{\ell^2}+\frac{16G^2J^2}{r^2}\right)\, ,
  \qquad N_\vphi=\frac{4GJ}{r^2}\, .                       \lab{2.6}
\eea
For the black hole with torsion, the triad field is taken as
\bsubeq\lab{2.7}
\be b^0=Ndt\, ,\qquad b^1=N^{-1}dr\, ,\qquad
b^2=r\left(d\vphi+N_\vphi dt\right)\, ,                    \lab{2.7a}
\ee
while the connection, in accordance with \eq{2.2}, has the form
\be
\om^i=\tom^i+\frac{p}{2}b^i\, ,                            \lab{2.7b}
\ee
where the Riemannian connection $\tom^i$ reads:
\be
\tom^0=-Nd\vphi\, ,\qquad \tom^1=N^{-1}N_\vphi dr\, ,\qquad
\tom^2= -\frac{r}{\ell}\frac{dt}{\ell}-rN_\vphi d\vphi\, . \lab{2.7c}
\ee
\esubeq

Energy and angular momentum of the black hole with torsion differ
from the corresponding GR expressions:
\be
E=m+\frac{\a_3}{a}\left(\frac{pm}{2}-\frac{J}{\ell^2}\right)\, ,
\qquad M=J+\frac{\a_3}{a}\left(\frac{pJ}{2}-m\right)\, .   \lab{2.8}
\ee

The black hole manifold is topologically $R^2\times S^1$. The AdS
solution (\ads3) is locally isometric to the black hole and can be
formally obtained from \eq{2.7} by taking $J=0$ and $8Gm=-1$.
However, \ads3\  has a different topology in which $\vphi$ is
\emph{not periodic}.

\prg{Supersymmetric extension.} There exists a simple locally
supersymmetric extension of 3D gravity with torsion, based on the
action \eq{2.4}. The extension includes two gravitini fields and is
usually referred to as $N=1+1$ AdS supergravity \cite{2,12,13}.
Consider the action
\be I=I_0
  -g\int\left(\bar\psi\nabla\psi-i\m\bar\psi b^i\g_i\psi\right)
  -g'\int\left(\bar\psi'\nabla\psi'
               -i\m'\bar\psi' b^i\g_i\psi'\right)\, ,      \lab{2.9}
\ee
where $\psi^\Pi$ are the gravitini fields (1-forms),
$\nabla\psi^\Pi=\bigl(d-\frac{i}{2}\om^m\g_m\bigr)\psi^\Pi$ are their
covariant derivatives, and Dirac matrices are pure imaginary. The
action is invariant under the local sypersymmetry transformations
with spinorial parameters $\ve^\Pi=(\ve,\ve')$:
\bea
&&\d_S b^i=i\bar\ve\g^i\psi +i\bar\ve'\g^i\psi'\, ,        \nn\\
&&\d_S\om^i=-2i\m'\bar\ve\g^i\psi-2i\m\bar\ve'\g^i\psi'\, ,\nn\\
&&\d_S\psi^\Pi=4a\left(
  \nabla\ve^\Pi-i\m^\Pi b^k\g_k\ve^\Pi\right)\, .          \lab{2.10}
\eea
provided the coupling constants $g,g'$ and $\m^\Pi=(\m,\m')$ satisfy
the relations
\bea
&&2ag=a-2\m'\a_3\, ,\qquad 2ag'=a-2\m\a_3\, ,              \nn\\
&&2\m+\frac{p}{2}=\frac{1}{\ell}\, ,\qquad
  2\m'+\frac{p}{2}=-\frac{1}{\ell}\, .                     \nn
\eea
Here, $\ell$ is the AdS radius, $\ell^{-1}=\m-\m'$, and for $\m$ and
$\m'$ real and different from each other, the effective cosmological
constant is negative: $\Leff=-(\m-\m')^2\equiv -{1}/{\ell^2}<0$.

The commutator algebra of the local super-Poincar\'e transformations
clo\-ses on shell, which is sufficient for exploring the asymptotic
symmetries.

The variation of the action with respect to $b^i$ and $\om^i$ yields
the supersymmetric extension of the field equations \eq{2.4}, the
variations with respect to $\bar\psi$ and $\bar\psi'$ yields the
gravitini field equations. The black hole and \ads3\ can be regarded
as exact solutions of the supersymmetric field equations with zero
gravitini, $\psi=\psi'=0$.

\section{Killing spinors} 
\setcounter{equation}{0}

Since the black hole and \ads3 are maximally symmetric solutions of
3D gravity, their symmetries are locally identical. However, the
existence of different global structures implies completely different
symmetries in the large.

\prg{Exact supersymmetries.} In the supersymmetric theory \eq{2.9},
the supersymmetry transformations that leave the black hole or \ads3\
configuration with $\psi=\psi'=0$ invariant, are called exact
supersymmetries. Since the conditions $\d_S b^i=\d_S\om^i=0$ are
automatically satisfied for $\psi=\psi'=0$, the spinor parameters
$(\ve,\ve')$ of exact supersymmetries are defined solely by the
requirements $\d_S\psi=\d_S\psi'=0$.
Expressing $\om^i$ in terms of the Levi-Civita connection $\tom^i$ as
in \eq{2.7b}, these requirements take the form
\be
\tilde\nabla\ve=\frac{i}{2\ell} b^k\g_k\ve\, ,\qquad
\tilde\nabla\ve'=-\frac{i}{2\ell} b^k\g_k\ve'\, ,          \lab{3.1}
\ee
where $\tilde\nabla$ is Riemannian covariant derivative. Equations
\eq{3.1} are called the Killing spinor equations; they define the
spinor parameters $(\ve,\ve')$ of the exact supersymmetries. For the
configuration \eq{2.7}, the Killing spinor equations take the form:
\bsubeq
\bea
&&\pd_+\ve=0\, ,\qquad
  \pd_-^2\ve-2G\left(m-\frac{J}{\ell}\right)\ve=0\, ,      \nn\\
&&\pd_1\ve+\frac {N^{-1}}2\left(\frac {4GJ}{r^2}
          +\frac{1}{\ell}\right)\s^3\ve=0\, ,              \lab{3.2}
\eea
and similarly:
\bea
&&\pd_-\ve'=0\, ,\qquad
  \pd_+^2\ve'-2G\left(m+\frac{J}{\ell}\right)\ve'=0\, ,    \nn\\
&&\pd_1\ve'+\frac{N^{-1}}2\left(\frac{4GJ}{r^2}
  -\frac{1}{\ell}\right)\s^3\ve'=0\, ,
\eea
\esubeq
where $x^\pm=t/\ell\pm\vphi$. These equations depend on the black
hole parameters $m$ and $J$ through the combination
$\l^2_\mp=2G\left(m\mp J/\ell\right)$. Among their solutions, only
those that are in agreement with the global properties of the black
hole/\ads3\ are acceptable.

\prg{Killing spinors for AdS\boldmath{$_3$}.} The AdS solution is
characterized by $\l^2_-=\l^2_+=-1/4$, and it possesses four Killing
spinors, two for $\ve$ and two for $\ve'$:
\bea
&&\ve=\left(\sqrt{\frac{N_*+1}{2}}\,\s^3
           -\sqrt{\frac{N_*-1}{2}}\,\right)
  \left(\cos\frac{x^-}2+i\s^2\sin\frac{x^-}2\,\right)\zeta\,,\nn\\
&&\ve'=\left(\sqrt{\frac{N_*+1}{2}}\,\s^3
            +\sqrt{\frac{N_*-1}{2}}\,\right)
  \left(\cos\frac{x^+}2+i\s^2\sin\frac{x^+}{2}\right)\zeta'\, .
\eea
Here, $N_*=\sqrt{1+r^2/\ell^2}$ is the value of $N$ at \ads3\ and
$\zeta,\zeta'$ are constant spinors. The Killing spinors are
anti-periodic, which is in agreement with the global structure of
\ads3.

\prg{Killing spinors for the black hole.} Since the black hole is
obtained from \ads3\ by a process of identification, it possesses
locally the same number of Killing spinors as \ads3. Globally,
however, spinors on the black hole manifold can be either periodic
or anti-periodic \cite{4,x1,15,x9}, which is a serious restriction on
the solutions of the Killing spinor equations \eq{3.2}. The form
of this restriction depends on the values of the black hole
parameters $m$ and $J$.

The reality of the black hole horizons, ensured by the condition
$|J|\le m\ell$, implies $\l^2_\mp\geq 0$.

\prg{1.} For a \emph{generic black hole} with $|J|<m\ell$, we have
$\l^2_->0$ and $\l^2_+>0$. Consequently, there are no
periodic/anti-periodic solutions of \eq{3.2}, and no Killing spinors.

\prg{2.} Consider now the \emph{extreme black hole} with $|J|=m\ell$
and $m\ne 0$.

(a) For $J=m\ell$, we have $\l^2_-=0$ and $\l^2_+>0$. The
periodic/anti-periodic solution for $\ve'$ does not exist, while the
solution for $\ve$ has the form
\bsubeq\lab{3.4}
\be
\ve=\left[\left(|u|^{1/2}+|u|^{-1/2}\right)-
    {\rm sgn}u\left(|u|^{1/2}-|u|^{-1/2}\right)\s_3\right]
    \frac{1-x^-\s^-}2\zeta\, ,
\ee
where $u=r/\ell-4Gm\ell/r$. The Killing spinor $\ve$ is compatible
with the periodicity in $\vphi$ if the term linear in $x^-$
dissapears, i.e. if
$$
\zeta\sim\left(\ba{c} 0 \\
                      1 \ea\right)\, .
$$
Thus, we have here only \emph{one} Killing spinor.

(b) Similarly, in the case $J=-m\ell$, i.e. $\l^2_->0$ and
$\l^2_+=0$, there are no periodic/anti-periodic solutions for $\ve$,
while globally acceptable solution for $\ve'$ takes the form
\be
\ve'=\left[\left(|u|^{1/2}+|u|^{-1/2}\right)
     +{\rm sgn}u\left(|u|^{1/2}-|u|^{-1/2}\right)\s_3\right]
      \frac{1}{2}\left(\ba{c} 1 \\
                              0 \ea\right)\, .
\ee
\esubeq

\prg{3.} The case $m=J=0$, with $\l^2_-=0=\l^2_+=0$, corresponds to
the \emph{zero-energy black hole}, for which both $E$ and $M$ vanish.
The related Killing spinors can be obtained from the extreme black
hole solutions \eq{3.4} in the limit $m\to 0$. The periodicity
requirement implies the existence of \emph{two} Killing spinors:
\bea
\ve\sim\sqrt{\frac{r}{\ell}}
   \left(\ba{c}
           0 \\
           1
         \ea\right)\, ,\qquad
\ve'\sim\sqrt{\frac{r}{\ell}}
   \left(\ba{c}
           1 \\
           0
         \ea\right)\, .                                    \lab{3.5}
\eea

Thus, \emph{a generic black hole has no Killing spinors, the extreme
black hole has only one, while the zero-energy black hole has two}.

\section{Stability} 
\setcounter{equation}{0}

The existence of Killing spinors for the black hole/\ads3, combined
with the asymptotic supersymmetry algebra, has serious implications
for the stability of these solutions.

\prg{Asymptotic symmetries.} For $\Leff<0$, the asymptotic
structure of 3D gravity with torsion is well understood \cite{10,x6}.
It can be generalized to the supersymmetric theory \eq{2.9} by a
suitable completion of the asymptotic conditions in the fermionic
sector. The asymptotic symmetry is defined as a subset of local
super-Poincar\'e transformations that leaves the adopted
asymptotic configuration invariant. It is characterized by four
chiral parameters: two of them are bosonic, $T^\mp(x^\mp)$, while
the other two are fermionic, $\eps^\mp(x^\mp)$. The asymptotic
Poisson bracket algebra of the canonical generators
$\tG(T^\mp,\eps^\mp)$ is given (in the quantum-mechanical
notation) by two independent copies of the super-Virasoro algebra
\cite{13}: \bea \left[L^\mp_n,L^\mp_m\right]&=&
   (n-m)L^\mp_{n+m}+\frac{c^\mp}{12}n^3\d_{m+n,0}\, ,      \nn\\
\left[L^\mp_n,Q^{\mp}_m\right]&=&
   \left(\frac{1}{2}n-m\right)Q^\mp_{m+n}\, ,              \nn\\
\left\{Q_n^\mp,Q_m^\mp\right\}&=&
   2L^\mp_{n+m}+\frac{c^\mp}{3}n^2\d_{m+n,0}\, ,           \lab{4.1}
\eea
where $L^\mp_n$ and $Q^\mp_n$ are Fourier modes of
$\tG(T^\mp,\eps^\mp)$, and $c^\mp$ are classical central charges:
$$
c^-=12\cdot 2\pi\ell ag\, ,\qquad c^+=12\cdot 2\pi\ell ag'\, .
$$
In the sectors with periodic (Ramond) or anti-periodic
(Neveau-Schwartz) boundary conditions for fermions \cite{4,x1}, index of
$Q^\mp_m$ takes on integer or half-integer values, respectively. The
reality properties of the modes are: $(L^\mp)^\dag_n=L^\mp_{-n}$,
$(Q^\mp)^\dag_n=Q^\mp_{-n}$.

The eigenvalues of the bosonic operators $L_0^\mp$ can be expressed
in terms of the energy $E$ and angular momentum $M$ of the system.
For the black hole/\ads3\ solution \eq{2.7}, we have:
\be
L_0^\mp=\frac{\ell E\pm M}{2}
       =(\ell m\pm J)\frac{c^\mp}{48\pi a\ell}\, .
\ee

One should stress that the asymptotic supersymmetry algebra \eq{4.1}
is defined by \emph{ignoring} (factoring out) pure gauge
transformations \cite{1,10,x6,15,x9}. The algebra has interesting
implications on the stability properties of the black hole/\ads3.

\prg{Ramond sector.} In the Ramond sector (with periodic boundary
conditions), the subalgebra of \eq{4.1} with vanishing central
charge, generated by $(L_0^\mp,Q_0^\mp)$, yields:
\be
L^\mp_0=Q^\mp_0Q^\mp_0=Q^\mp_0(Q^\mp_0)^\dag\ge 0\, .      \lab{4.3}
\ee
These relations are equivalent to $|M|\le \ell E$, which implies
$E\ge 0$.

We have seen that the zero-energy black hole configuration has two
Killing spinors, which are the parameters of two exact
supersymmetries, generated by the action of $Q_0^\mp$. On the other
hand, the zero-energy black hole \emph{saturates} the bounds
\eq{4.3}, which means that it has the lowest values of the conserved
charges $L^\mp_0$, compared to any other black hole configuration.
Consequently, \emph{the zero-energy black hole can be naturally
interpreted as the black hole ground state} (the ground state of the
Ramond sector).

One should note that for an extremal black hole, only one of the
bounds in \eq{4.3} is saturated, while for a generic black hole, both
$L^-_0$ and $L^+_0$ are strictly positive.

\prg{Neveu-Schwarz sector.} The AdS solution belongs to the
Neveu-Schwarz sector (anti-periodic boundary conditions), with (the
value of $L^\mp_0$) = $-c^\mp/24$. If one makes the shift
$L_0^\pm\rightarrow L_0^\pm-c^\pm/24$, the value of the new $L^\mp_0$
on \ads3\ vanish. The $osp(1|2)\otimes osp(1|2)$ subalgebra with
vanishing charge is generated by $(L_{\mp 1}, L_0,Q_{\mp 1/2})^\mp$,
and it implies
\be
2L^\mp_0=Q^\mp_{1/2}Q^\mp_{-1/2}+Q^\mp_{-1/2}Q^\mp_{1/2}
        =Q^\mp_{1/2}(Q^\mp_{1/2})^\dag
         +Q^\mp_{-1/2}(Q^\mp_{-1/2})^\dag\ge 0\, .         \lab{4.4}
\ee

The AdS solution has four Killing spinors, which are the parameters
of four exact supersymmetries, generated by the action of
$(Q^\mp_{1/2},Q^\mp_{-1/2})$, and it saturates the bounds \eq{4.4}.
Thus, \emph{we are led to interpret AdS$_3$ as the ground state of the
Neveu-Schwarz sector}.

\section{Concluding remarks} 

In this paper, we used the $N=1+1$ sypersymmetric extension of 3D
gravity with torsion to investigate exact supersymmetries and stability
of the black hole with torsion.

(1) We showed that a generic black hole has no exact supersymmetries,
the extremal black hole has only one, while the zero-energy black
hole has two exact supersymmetries.

(2) Using the asymptotic superconformal symmetry of the theory, we
found that the zero-energy black hole and \ads3\ can be naturally
interpreted as the ground states of the sectors with periodic
and anti-periodic boundary conditions, respectively.

These results are a natural generalization of those found earlier in
Riemannian GR.

\section*{Acknowledgments} 

This work was supported by the Serbian Science Foundation.

\appendix

\end{document}